\documentclass[conference]{IEEEtran}
\IEEEoverridecommandlockouts
\usepackage{cite}
\usepackage{amsmath,amssymb,amsfonts}
\usepackage{algorithmic}
\usepackage{graphicx}
\usepackage{textcomp}
\usepackage{xcolor}

\usepackage{algorithm}
\usepackage{array}
\usepackage{makecell}
\usepackage[caption=false,font=footnotesize,labelfont=sf,textfont=sf]{subfig}
\usepackage{textcomp}
\usepackage{stfloats}
\usepackage{url}
\usepackage{verbatim}
\usepackage{graphicx}
\usepackage{cite}
\usepackage{multirow}
\usepackage{multicol}
\usepackage{tabularx}
\usepackage{array} % for >{\centering\arraybackslash}X
\usepackage{multirow}
\usepackage{booktabs}
\usepackage{adjustbox}
\usepackage{cleveref}
\usepackage[table]{xcolor}
\usepackage{pifont}

\def\BibTeX{{\rm B\kern-.05em{\sc i\kern-.025em b}\kern-.08em
    T\kern-.1667em\lower.7ex\hbox{E}\kern-.125emX}}
\begin{document}

\title{AudioGS: Spectrogram-Based Audio Gaussian Splatting for Sound Field Reconstruction
\thanks{\textsuperscript{*}Corresponding authors. This work was partly supported by the NSFC (62431015, 62571317, 62501387), the Fundamental Research Funds for the Central Universities, Shanghai Key Laboratory of Digital Media Processing and Transmission under Grant 22DZ2229005, Special Fund for Promoting High-Quality Industrial Development (2025358) and the 111 Project BP0719010.}
}

\author{
    \IEEEauthorblockN{
        Chunhao Bi\textsuperscript{1}, 
        Houqiang Zhong\textsuperscript{1}, 
        Zhixin Xu\textsuperscript{2}, 
        Li Song\textsuperscript{1*}, 
        Zhengxue Cheng\textsuperscript{1*}
    }
    \IEEEauthorblockA{
        \textsuperscript{1}\textit{School of Electronic Information and Electrical Engineering, Shanghai Jiao Tong University}, Shanghai, China \\
        \textsuperscript{2}\textit{Institute of Cultural and Creative Industry, Shanghai Jiao Tong University}, Shanghai, China \\
        Emails: \{bichunhao, zhonghouqiang, zhixin.xu, song\_li, zxcheng\}@sjtu.edu.cn
    }
}
\maketitle

\begin{abstract}
Spatial audio is fundamental to immersive virtual experiences, yet synthesizing high-fidelity binaural audio from sparse observations remains a significant challenge.
Existing methods typically rely on implicit neural representations conditioned on visual priors, which often struggle to capture fine-grained acoustic structures. 
Inspired by 3D Gaussian Splatting (3DGS), we introduce AudioGS, a novel visual-free framework that explicitly encodes the sound field as a set of Audio Gaussians based on spectrograms.
AudioGS associates each time-frequency bin with an Audio Gaussian equipped with dual spherical harmonic (SH) coefficients and a decay coefficient.
For a target pose, we render binaural audio by evaluating the SH field to capture directionality, incorporating geometry-guided distance attenuation and phase correction, and reconstructing the waveform.
Experiments on the Replay-NVAS dataset demonstrate that AudioGS successfully captures complex spatial cues and outperforms state-of-the-art visual-dependent baselines.
Specifically, AudioGS reduces the magnitude reconstruction error (MAG) by over \textbf{14\%} and reduces the perceptual quality metric (DPAM) by approximately \textbf{25\%} compared to the best performing visual-guided method.
\end{abstract}


\section{Introduction} \label{sec:intro}
Recent advancements in immersive technologies, such as VR, AR, and XR, have democratized access to 3D content. Originally rooted in gaming, these applications have rapidly expanded into diverse domains, including education, personal communication, and virtual conferencing.
Spatial audio is important for these experiences, enhancing realism by simulating a physical auditory space. However, despite the progress in visual rendering, audio often remains limited to conventional channel-based formats. These inputs lack genuine spatial depth, creating a sensory mismatch—where sound is perceived as ``flat'' regardless of user movement—that severely undermines immersive realism.

To address this challenge, Chen et al.~\cite{chen2023novel} introduced the task of novel-view acoustic synthesis (NVAS). NVAS aims to synthesize binaural audio from a novel listener pose, given visual and acoustic input from another source viewpoint in the same scene. 
Existing methods, such as ViGAS~\cite{chen2023novel} and AV-NeRF~\cite{liang2023av}, typically employ implicit neural representations conditioned on visual priors.
However, relying on learned cross-modal priors to predict the acoustic environment often fails to capture the unique, fine-grained details of the sound field, leading to over-smoothed spectral estimations.
Additionally, these methods neglect the modeling of phase shifts by simply reusing the source phase for the target view, resulting in inaccurate Inter-Aural Time Difference (ITD) cues and significant spatial misalignment.
This raises a natural question: \textbf{Is it possible to explicitly reconstruct the 3D sound field solely from sparse acoustic observations, and use it to synthesize high‑fidelity binaural audio at unseen listener poses without relying on visual priors?}

\begin{figure}[t]
\centering
\includegraphics[width=\linewidth]{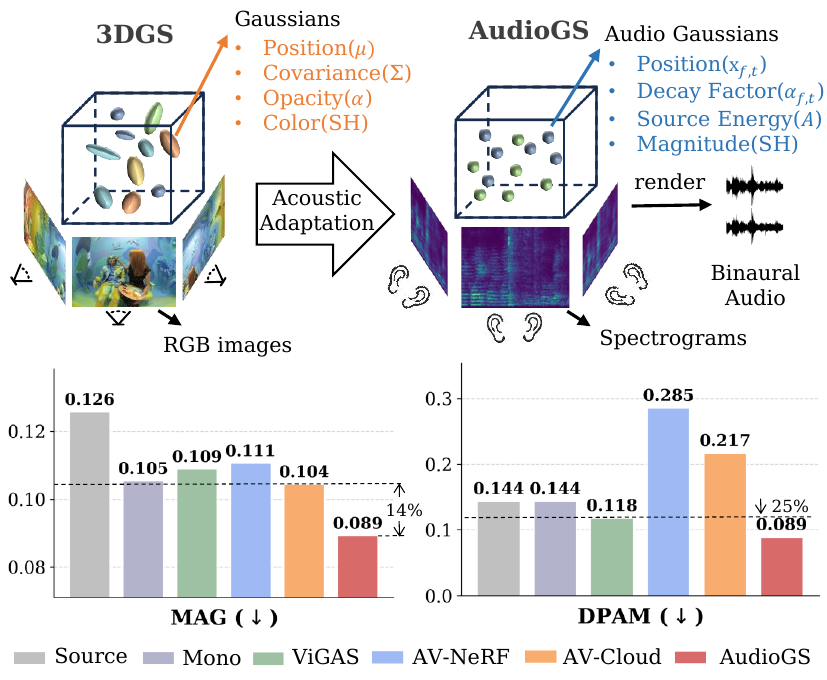}
\caption{\textbf{AudioGS framework and performance.}
\textbf{Top:} Inspired by 3D Visual Gaussian Splatting, we propose AudioGS, a visual-free framework that models the sound field as a set of learnable 3D Audio Gaussians.
\textbf{Bottom:} Quantitative comparison on the Replay-NVAS dataset. AudioGS significantly outperforms baselines in reconstruction quality (MAG) and perceptual fidelity (DPAM).}
\label{fig:teaser}
\end{figure}

Inspired by the great success of 3D Gaussian Splatting (3DGS)~\cite{kerbl20233d} in novel view synthesis, we propose AudioGS, a novel visual-free framework that explicitly models the sound field using a set of learnable \textbf{Audio Gaussians} without relying on visual priors. 
We adapt 3DGS by replacing anisotropic covariance with a physical decay coefficient, formulating Audio Gaussians as isotropic radiators governed by wave propagation. Specifically, as Fig.~\ref{fig:teaser}, each Gaussian corresponds to a spectrogram bin and utilizes Spherical Harmonics (SH) to encode directional energy.
Our main contributions are summarized as follows: 
1) We introduce a visual-free explicit representation for the NVAS task. We formulate the sound field as a set of Audio Gaussians directly mapped from specific time-frequency bins. This explicit representation allows for compact and interpretable modeling of the sound field.
2) We design a dual SH parameterization to decouple directional energy from spatial binaural cues, integrated with geometry-guided distance decay and phase correction.
3) Experiments on the Replay-NVAS dataset demonstrate that AudioGS significantly outperforms existing state-of-the-art visual-dependent methods. As shown in Fig.~\ref{fig:teaser}, without visual data, our method reduces the magnitude reconstruction error (MAG) by over \textbf{14\%} and improves perceptual quality (DPAM) by approximately \textbf{25\%} compared to the visual-guided SOTA baselines.

\begin{figure*}[tb]
\centerline{\includegraphics[width=\linewidth]{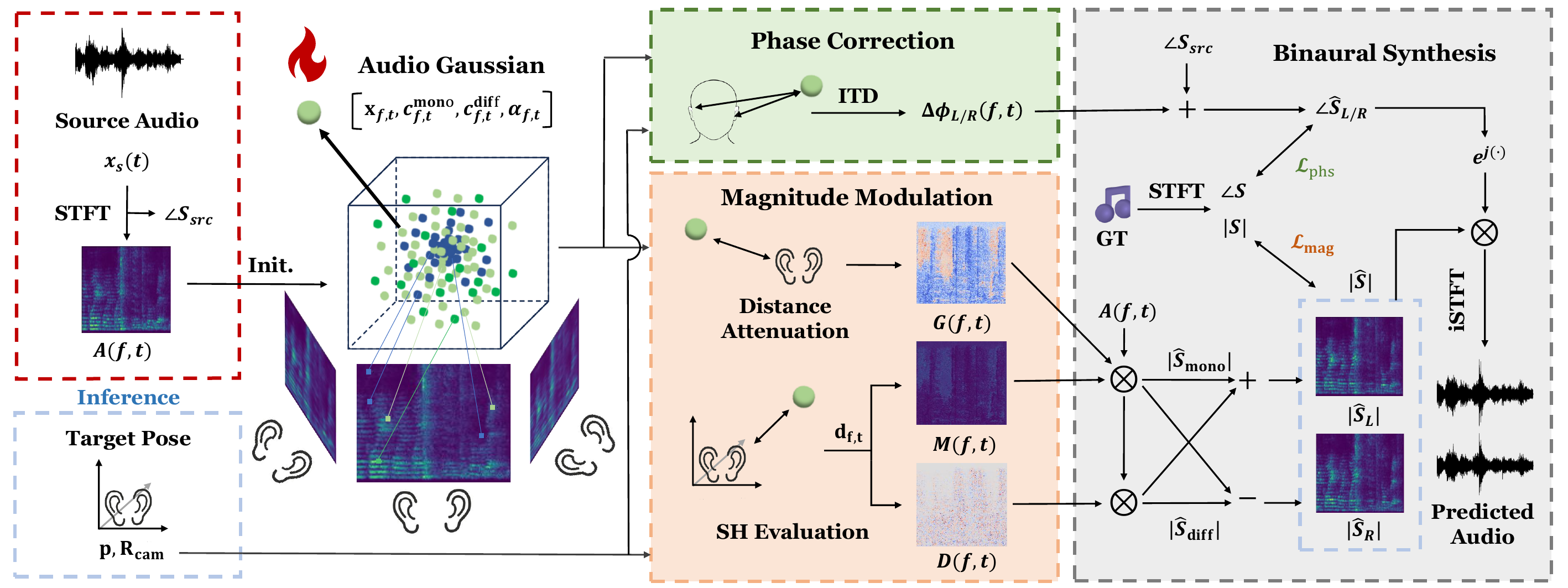}}
\caption{\textbf{Overview of the AudioGS framework.} 
We explicitly represent the sound field by encoding the source spectrogram into a set of Audio Gaussians, where each Gaussian corresponds to a specific time-frequency bin. 
For a target listener pose, the rendering pipeline separates into two streams: 
(1) \textbf{Magnitude Modulation}, which utilizes spherical harmonics and distance attenuation to model directional energy and spatial decay; and
(2) \textbf{Phase Correction}, which computes inter-aural time differences (ITD) based on wave propagation delays.
Finally, the modulated magnitudes and corrected phases are combined via inverse STFT (iSTFT) to synthesize high-fidelity binaural audio.}
\label{fig:overview}
\end{figure*}

\section{Related Work} \label{sec:related work}

\subsection{Acoustic Modeling and Field Representation}

Traditional acoustic modeling typically relies on the convolution of source audio with RIR. Early approaches are based either on numerically solving the wave equation~\cite{savioja2015overview} or on the assumptions of geometrical acoustics~\cite{tang2020improving}. Recent methods leverage deep learning to synthesize spatial RIR~\cite{ratnarajah2022fast}.
However, these RIR generators typically require source positions and detailed scene geometry, which are often challenging to measure in real-world scenarios. 
To bypass geometric constraints, researchers have proposed reconstructing spatial audio directly from visual imagery. Chen et al.~\cite{chen2022visual} proposed a cross-modal Transformer to infuse visual attributes into the audio stream via attention mechanisms. Similarly, Mono2Binaural~\cite{gao20192} utilizes a CNN to predict binaural signals consistent with the visual spatial layout, while PseudoBinaural~\cite{xu2021visually} uses spherical harmonic decomposition to render spatial audio mapped from video pixel coordinates. However, these works only synthesize acoustics for a given viewpoint rather than a novel viewpoint.

\subsection{Novel-View Acoustic Synthesis}
In order to synthesize the sound in a scene from a novel viewpoint, Chen et al.~\cite{chen2023novel} introduced the NVAS task and ViGAS, which learns to synthesize spatial audio guided by visual observations. Subsequently, AV-NeRF~\cite{liang2023av} integrated acoustic rendering into Neural Radiance Fields (NeRF) to jointly synthesize novel-view video and audio, though it remains restricted to static, single-source scenes. Most recently, AV-Cloud~\cite{chen2024av} adopted an explicit point-based approach, employing Structure-from-Motion (SfM) to reconstruct sparse audio-visual anchors for geometry modeling. However, AV-Cloud heavily relies on accurate camera calibration and visual textures to obtain SfM points, causing performance degradation in untextured or noisy environments.

\subsection{3D Gaussian Splatting for Acoustic Tasks}
3DGS explicitly models 3D scenes using anisotropic 3D Gaussians optimized via differentiable rendering. While successful in vision, its acoustic applications are limited. Reference~\cite{yoshida2025extending} encodes audio into points with SH coefficients but relies on a U-Net for rendering, diverging from the explicit splatting pipeline. AV-GS~\cite{bhosale2024av} integrates audio-guided parameters into 3DGS to capture holistic scene conditions and geometry. However, similar to AV-Cloud, AV-GS depends heavily on visual reconstruction quality. Consequently, its performance suffers in scenarios with sparse visual data or inconsistent camera calibration.
Thus, we thus propose AudioGS, a spectrogram-based Gaussian Splatting framework to represent a 3D sound field without relying on visual priors.

\section{Methods} \label{sec:methods}

AudioGS introduces a novel explicit representation for sound field reconstruction. As illustrated in Fig.~\ref{fig:overview}, our framework directly encodes the audio spectrogram into a set of learnable 3D Audio Gaussians, without using visual priors.

\subsection{3DGS Modeling of Spatial Audio}
\label{subsec:modeling}
Given a training sample consisting of a source binaural waveform $x_s(t)$ recorded at a reference listener position $\mathbf{p}_{\mathrm{ref}} \in \mathbb{R}^3$, a target listener pose $(\mathbf{p}, R_{\mathrm{cam}})$, and the
corresponding ground-truth binaural waveform $y(t)$ at the target, we aim to synthesize the spatial audio at $(\mathbf{p}, R_{\mathrm{cam}})$.
We transform $x_s(t)$ into the time-frequency domain via STFT, obtaining complex spectrograms
$S_{\mathrm{src},k}(f,t)\in\mathbb{C}^{F\times T}$ for $k\in\{L,R\}$, where $F$ and $T$ denote the numbers of frequency bins and time frames respectively.
We define the source content magnitude as $A(f,t)=\frac{1}{2}\left(|S_{\mathrm{src},L}(f,t)|+|S_{\mathrm{src},R}(f,t)|\right)$, and keep the per-ear source phases $\angle S_{\mathrm{src},k}(f,t)$ for reconstruction.

Our explicit representation leverages the sparsity of speech signals in the T-F domain.
Jourjine et al.~\cite{jourjine2000blind} showed that speech mixtures exhibit approximate W-disjoint orthogonality, suggesting that the energy at a given T-F bin is often dominated by a single source component.
Furthermore, in the context of SH-domain processing, Cobos et al.~\cite{cobos2023acoustic} demonstrated that the sound-field coefficients at a specific $(f,t)$ bin can be factorized into a source-dependent term and a direction-dependent term (Modal Directional Pattern, MDP).
Together, these observations motivate mapping each T-F bin to an Audio Gaussian as an explicit spatial primitive.

\noindent\textbf{AudioGS Initialization.}
We associate each STFT bin $(f,t)$ with an Audio Gaussian and represent the sound field as
\begin{equation}
\mathcal{G}=\{(\mathbf{x}_{f,t},\mathbf{c}^{\mathrm{mono}}_{f,t},\mathbf{c}^{\mathrm{diff}}_{f,t},\alpha_{f,t})\}.
\end{equation}
Here $(f,t)$ indexes a bin on the $F\times T$ STFT grid (i.e., $f\in\{1,\dots,F\}$ and $t\in\{1,\dots,T\}$).
Each Audio Gaussian is parameterized by a learnable 3D position $\mathbf{x}_{f,t}\in\mathbb{R}^3$, dual SH coefficient vectors $\mathbf{c}^{\mathrm{mono}}_{f,t}$ and $\mathbf{c}^{\mathrm{diff}}_{f,t}$, and a learnable distance decay coefficient $\alpha_{f,t}>0$.
To decouple acoustic content from spatial geometry, we treat $A(f,t)$ as the source energy carrier and use SH coefficients purely for directional modulation: $\mathbf{c}^{\mathrm{mono}}_{f,t}$ produces a non-negative shared (monaural) energy pattern, while $\mathbf{c}^{\mathrm{diff}}_{f,t}$ encodes signed left-right difference cues for binaural perception.

\noindent\textbf{Magnitude Modulation.}
To render the audio for a listener at position $\mathbf{p}$ with orientation $R_{\mathrm{cam}}$, we first calculate the relative direction $\mathbf{d}_{f,t}$ from each Gaussian $\mathbf{x}_{f,t}$ to the listener. We then employ spherical harmonics to model the directional energy distribution. The mono mask $M(f,t)$ and difference mask $D(f,t)$ are obtained by projecting the learned coefficients onto the SH basis functions $\mathbf{Y}(\mathbf{d}_{f,t})$~\cite{zhang2022differentiable}:
\begin{equation}
\begin{aligned}
M(f,t) &= 2 \cdot \mathrm{sigmoid}\left( \langle \mathbf{c}^{\mathrm{mono}}_{f,t}, \mathbf{Y}(\mathbf{d}_{f,t}) \rangle \right), \\
D(f,t) &= \langle \mathbf{c}^{\mathrm{diff}}_{f,t}, \mathbf{Y}(\mathbf{d}_{f,t}) \rangle.
\end{aligned}
\end{equation}

To adhere to physical wave propagation laws~\cite{bradley1986predictors}, we introduce a distance attenuation term. We model the spectral magnitude decay with respect to a reference training pose $\mathbf{p}_{\mathrm{ref}} \in \mathbb{R}^3$:
\begin{equation}
G(f,t) = \left( \frac{\| \mathbf{p}_{\mathrm{ref}} - \mathbf{x}_{f,t} \|_2 + \varepsilon}{\| \mathbf{p} - \mathbf{x}_{f,t} \|_2 + \varepsilon} \right)^{\alpha_{f,t}}.
\end{equation}
where $\varepsilon$ is a numerical stability term and $\alpha_{f,t} > 0$ is a learnable decay coefficient. This allows the model to adaptively capture varying acoustic propagation characteristics, ranging from reverberant fields ($\alpha$ close to $0$) to direct path attenuation ($\alpha \approx 1$).

\noindent\textbf{Phase Correction.}
\label{subsec:phase}
Let $\theta_{f,t}$ denote the signed azimuth of $\mathbf{x}_{f,t}$ in the listener frame.
We compute a non-negative ITD magnitude $\tau_{\text{phys}}(f,|\theta|)$ using a rigid sphere
model~\cite{aaronson2014testing}, and introduce a bounded multiplicative residual $\eta_{f,t}=1+
\lambda\tanh(\delta_{f,t})$. $\lambda$ controls the residual range.
Denoting the reference-view azimuth as $\theta^{\mathrm{ref}}_{f,t}$, the per-ear phase correction is:
\begin{equation}
\Delta \phi_{k}(f,t)=\mp \frac{\omega_f}{2}\,\eta_{f,t}\Big(\tau_{\text{phys}}(f,|\theta_{f,t}|)-\tau_{\text{phys}}
(f,|\theta^{\mathrm{ref}}_{f,t}|)\Big),
\end{equation}
where $\mp$ is negative/positive for the contralateral/ipsilateral ear, and $\omega_f$ is the angular frequency of STFT bin $f$.

\subsection{Binaural Synthesis}
\label{subsec:synthesis}

To avoid shortcut learning in spatialization~\cite{gao20192}, we adopt a cascaded strategy where the difference envelope modulates the predicted mono energy.
We first compute the mono magnitude $|\hat{S}_{\mathrm{mono}}|$ by applying distance attenuation $G(f,t)$ and the mono mask $M(f,t)$ to the source content $A(f,t)$.
Subsequently, the Left ($L$) and Right ($R$) magnitudes are derived by adding or subtracting the difference component, which is scaled relative to the mono energy:
\begin{equation}
\begin{aligned}
    |\hat{S}_{\mathrm{mono}}(f,t)| &= A(f,t) \cdot G(f,t) \cdot M(f,t), \\
    |\hat{S}_{L/R}(f,t)| &= \mathrm{ReLU}\Big( |\hat{S}_{\mathrm{mono}}(f,t)| \cdot \left( 1 \pm D(f,t) \right) \Big).
\end{aligned}
\label{eq:mag_synthesis}
\end{equation}
Here, the term $|\hat{S}_{\mathrm{mono}}| \cdot D(f,t)$ represents the difference magnitude $|\hat{S}_{\mathrm{diff}}|$. The sign $\pm$ is positive for the left ear and negative for the right.
We combine these magnitudes with the corrected phase to reconstruct the complex spectrograms:
\begin{equation}
\hat{S}_k(f,t) = |\hat{S}_k(f,t)| \cdot \exp\left( j \left(\angle S_{\mathrm{src},k}(f,t) + \Delta
\phi_k(f,t)\right) \right),
\label{eq:binaural_final}
\end{equation}
for $k \in \{L, R\}$. Finally, the complex spectrograms are transformed back to the time domain via inverse STFT (iSTFT) to produce the binaural waveform.

\subsection{Training Loss}

Given the predicted binaural complex spectrograms $\hat{S}_{k}$ (from Eq.~\eqref{eq:binaural_final}) and the ground-truth $S_{k}$ for $k \in \{L, R\}$, we first transform them into mono and difference components:
\begin{equation}
    S_{\mathrm{mono}} = S_L + S_R, \quad S_{\mathrm{diff}} = S_L - S_R.
\end{equation}
We apply the same transformation to the predicted signals to obtain $\hat{S}_{\mathrm{mono}}$ and $\hat{S}_{\mathrm{diff}}$.

\noindent\textbf{Magnitude Loss.}
We minimize the L2 distance between the log-magnitude spectrograms~\cite{chen2023learning}:
\begin{equation}
    \mathcal{L}_{\mathrm{mag}}(S, \hat{S}) = \big\lVert \log|S| - \log|\hat{S}| \big\rVert_2^2.
\end{equation}

\noindent\textbf{Phase Loss.}
To address the discontinuity issues of phase wraparound, we map the phase angles to their corresponding rectangular coordinates on the unit circle:
\begin{equation}
    \mathcal{L}_{\mathrm{phs}}(S, \hat{S}) = \big\lVert \sin(\angle S) - \sin(\angle \hat{S}) \big\rVert_2^2 + \big\lVert \cos(\angle S) - \cos(\angle \hat{S}) \big\rVert_2^2,
\end{equation}
where $\angle S$ and $\angle \hat{S}$ denote the phase angles of the ground-truth and predicted spectrograms respectively.

\noindent\textbf{Total Loss.}
For brevity, we let $\mathcal{L}_{*}^{k}$ denote the loss term $\mathcal{L}_{*}(S_k, \hat{S}_k)$ for a component $k \in \{\mathrm{mono}, \mathrm{diff}, \mathrm{L}, \mathrm{R}\}$. The final objective aggregates the magnitude and phase losses for both mono and difference components:
\begin{equation}
    \mathcal{L}_{\mathrm{total}} = \mathcal{L}_{\mathrm{mag}}^{\mathrm{mono}} + \lambda_{\mathrm{diff}}\mathcal{L}_{\mathrm{mag}}^{\mathrm{diff}} + \lambda_{\mathrm{phs}} \left( \mathcal{L}_{\mathrm{phs}}^{\mathrm{L}} + \mathcal{L}_{\mathrm{phs}}^{\mathrm{R}} \right).
\end{equation}
where $\lambda_{\mathrm{diff}}$ and $\lambda_{\mathrm{phs}}$ are weighting factors for the spatial difference component and phase losses.

\newcommand{\cmark}{\ding{51}}%
\newcommand{\xmark}{\ding{55}}%

% 粉色
\definecolor{best}{HTML}{FF9999} 
% 橙黄色
\definecolor{second}{HTML}{FFE4B5} 
% 黄色
\definecolor{third}{HTML}{FFFFCC} 

\begin{table*}[t]
\centering
\caption{\textsc{Results on Replay-NVAS dataset.} 
Background colors indicate performance ranking: \colorbox{best}{\textbf{First}}, \colorbox{second}{\textbf{Second}}, and \colorbox{third}{\textbf{Third}} best results.}
\label{tab:comparison}

\resizebox{\textwidth}{!}{%
\begin{tabular}{l|c|cccc|cccc|cccc}
\hline
\multirow{2}{*}{Method} & \multirow{2}{*}{Visual} & \multicolumn{4}{c|}{\texttt{SC-1044}} & \multicolumn{4}{c|}{\texttt{SC-1052}} & \multicolumn{4}{c}{\texttt{SC-1074}} \\
 & & MAG$\downarrow$ & ENV$\downarrow$ & LRE$\downarrow$ & DPAM$\downarrow$ & MAG$\downarrow$ & ENV$\downarrow$ & LRE$\downarrow$ & DPAM$\downarrow$ & MAG$\downarrow$ & ENV$\downarrow$ & LRE$\downarrow$ & DPAM$\downarrow$  \\ \hline

Source Binaural & \xmark & 0.2468 & 0.1030 & 2.5450 & \cellcolor{third}0.1279
& 0.0966 & 0.0318 & 3.0069 & \cellcolor{third}0.1477
& 0.0547 & 0.0187 & 1.9118 & \cellcolor{third}0.1247 \\

Mono & \xmark & 0.2029 & 0.0836 & \cellcolor{third}0.5400 & 0.1280
& \cellcolor{third}0.0836 & 0.0269 & 0.8100 & \cellcolor{third}0.1477
& 0.0460 & 0.0158 & 0.5287 & \cellcolor{third}0.1247 \\

ViGAS~\cite{chen2023novel} & \cmark & \cellcolor{second}0.1945 & \cellcolor{best}0.0608 & 0.5517 & \cellcolor{second}0.1105 
& 0.0842 & \cellcolor{third}0.0225 & \cellcolor{best}0.4802 & \cellcolor{second}0.1062
& 0.0511 & 0.0147 & \cellcolor{second}0.4700 & \cellcolor{second}0.1095 \\

AV-NeRF~\cite{liang2023av} & \cmark & 0.2354 & 0.0942 & 0.6315 & 0.2278 
& 0.0883 & 0.0289 & 0.8424 & 0.2946
& \cellcolor{third}0.0456 & \cellcolor{second}0.0142 & 0.5770 & 0.4999 \\

AV-Cloud~\cite{chen2024av} & \cmark & \cellcolor{third}0.2004 & \cellcolor{third}0.0707 & \cellcolor{best}0.4380 & 0.2271 
& \cellcolor{second}0.0665 & \cellcolor{second}0.0204 & \cellcolor{second}0.6658 & 0.1621
& \cellcolor{second}0.0439 & \cellcolor{third}0.0143 & \cellcolor{best}0.4357 & 0.2546 \\

Ours & \xmark & \cellcolor{best}0.1720 & \cellcolor{second}0.0633 & \cellcolor{second}0.4496 & \cellcolor{best}0.0767
& \cellcolor{best}0.0609 & \cellcolor{best}0.0201 & \cellcolor{third}0.7521 & \cellcolor{best}0.0621
& \cellcolor{best}0.0428 & \cellcolor{best}0.0137 & \cellcolor{third}0.5247 & \cellcolor{best}0.0829 \\
\hline

% 下半部分表格：SC-1084, SC-1107, Overall
% 这里的 multicolumn{1} 之后加了一个空列占位符 & \multicolumn{1}{c|}{} 对应 Visual 列
\multicolumn{1}{l|}{} & \multicolumn{1}{c|}{} & \multicolumn{4}{c|}{\texttt{SC-1084}} & \multicolumn{4}{c|}{\texttt{SC-1107}} & \multicolumn{4}{c}{\texttt{Overall}} \\ 
 & & MAG$\downarrow$ & ENV$\downarrow$ & LRE$\downarrow$ & DPAM$\downarrow$ & MAG$\downarrow$ & ENV$\downarrow$ & LRE$\downarrow$ & DPAM$\downarrow$ & MAG$\downarrow$ & ENV$\downarrow$ & LRE$\downarrow$ & DPAM$\downarrow$  \\ \hline

Source Binaural & \xmark & 0.0507 & 0.0198 & 2.4710 & 0.1607 
& 0.1798 & 0.0617 & 2.5849 & \cellcolor{third}0.1569 
& 0.1257 & 0.0470 & 2.5039 & \cellcolor{third}0.1436 \\

Mono & \xmark & \cellcolor{third}0.0431 & 0.0178 & 0.4970 & 0.1607 
& \cellcolor{third}0.1512 & 0.0510 & 0.6053 & \cellcolor{third}0.1569 
& \cellcolor{third}0.1054 & 0.0390 & 0.5962 & \cellcolor{third}0.1436 \\

ViGAS & \cmark & 0.0439 & \cellcolor{third}0.0137 & \cellcolor{third}0.4949 & \cellcolor{third}0.1453 
& 0.1706 & \cellcolor{third}0.0474 & 0.5716 & \cellcolor{best}0.1179 
& 0.1089 & \cellcolor{second}0.0318 & \cellcolor{best}0.5137 & \cellcolor{second}0.1179 \\

AV-NeRF & \cmark & \cellcolor{best}0.0359 & \cellcolor{second}0.0128 & \cellcolor{best}0.4082 & \cellcolor{second}0.1245 
& \cellcolor{second}0.1477 & \cellcolor{second}0.0471 & \cellcolor{best}0.3531 & 0.2803 
& 0.1106 & 0.0394 & 0.5624 & 0.2854 \\

AV-Cloud & \cmark & 0.0474 & 0.0178 & 0.6211 & 0.2096 
& 0.1636 & 0.0528 & \cellcolor{third}0.5257 & 0.2306 
& \cellcolor{second}0.1044 & \cellcolor{third}0.0352 & \cellcolor{third}0.5373 & 0.2168 \\

Ours & \xmark & \cellcolor{second}0.0370 & \cellcolor{best}0.0122 & \cellcolor{second}0.4757 & \cellcolor{best}0.0846 
& \cellcolor{best}0.1336 & \cellcolor{best}0.0407 & \cellcolor{second}0.4325 & \cellcolor{second}0.1370 
& \cellcolor{best}0.0893 & \cellcolor{best}0.0300 & \cellcolor{second}0.5269 & \cellcolor{best}0.0887 \\
\hline
\end{tabular}%
}
\end{table*}

\section{Experiments} \label{sec:experiments}
\subsection{Experimental Setup}

\noindent\textbf{Dataset.}
% \paragraph{Dataset}
We evaluate our method on the \textit{Replay-NVAS} dataset~\cite{shapovalov2023replaymultimodalmultiviewacted,chen2023novel}, which captures synchronized multi-view RGB videos and spatial audio in real-world indoor environments. The dataset features diverse daily scenarios (e.g., conversations, dining) recorded from eight Digital Single Lens Reflex (DSLR) cameras arranged around a central region. 
We select five representative scenes with high-quality audio and camera calibration, processing each scene into 3-second audio clips for training and evaluation.
We utilize 7 of the 8 listener poses as the training set to optimize the representation, while the remaining pose is completely held out for testing.
This setting challenges the model to interpolate the sound field from sparse discrete samples rather than dense arrays.
Following ViGAS~\cite{chen2023novel}, we resample all audio to $16$~kHz and apply a $150$~Hz highpass filter to reduce low-frequency noise. In total, we utilize 660 seconds of audio data for evaluation.

\noindent\textbf{Implementation Details.}
We compute the STFT using a $512$-point FFT, a window length of $400$, a hop size of $160$ and a Hamming window.
For initialization, the 3D coordinates of the Audio Gaussians are initialized near the mean center of the DSLR cameras with a small radius. All SH coefficients are initialized to $0$ and the decay coefficients are initialized to $1$. We train for 60 epochs using Adam with separate learning rates for positions and the remaining parameters.

\noindent\textbf{Baselines.}
We compare AudioGS against the following methods:
1) \textbf{Source Binaural}: The unprocessed binaural audio at the source viewpoint.
2) \textbf{Mono}: Simply averaging the left and right channels of the source binaural audio.
3) \textbf{ViGAS}~\cite{chen2023novel}: A neural rendering method that achieves novel-view acoustic synthesis by analyzing audio-visual cues from source viewpoints.
4) \textbf{AV-NeRF}~\cite{liang2023av}: A NeRF-based system that synthesizes binaural audio for a given camera pose by first rendering a pair of RGB and depth images from the same camera position to guide the acoustic synthesis.
5) \textbf{AV-Cloud}~\cite{chen2024av}: An explicit point-based approach that adopts SfM to reconstruct sparse audio-visual anchors for geometry modeling. 
We use the official pre-trained model for ViGAS.
For AV-NeRF and AV-Cloud, we use the official implementations and train a separate model for each scene under the same train/test split.
AV-NeRF additionally requires a sound source position as input; since Replay-NVAS does not provide source annotations, we set it to the mean center of the DSLR cameras for each scene.

\subsection{Evaluation Metrics}
We report four quantitative metrics to assess reconstruction quality, spatial accuracy, and perceptual fidelity:
1) Magnitude Spectrogram Distance~\cite{chen2023novel} \textbf{(MAG) $\downarrow$}: The average L1 distance between the predicted and ground-truth magnitude spectrograms, measuring the audio quality in the time-frequency domain.
2) Envelope Distance~\cite{liang2023av} \textbf{(ENV) $\downarrow$}: The Euclidean distance between the Hilbert envelopes of the predicted and ground-truth waveforms, assessing the audio quality in the time domain.
3) Left-Right Energy Ratio Error~\cite{chen2023novel} \textbf{(LRE) $\downarrow$}: The absolute difference in the left-to-right energy ratio (in dB), evaluating the accuracy of spatial sound.
4) Deep Perceptual Audio Metric~\cite{manocha2020differentiable} \textbf{(DPAM) $\downarrow$}: A deep learning-based perceptual metric correlated with human judgment of audio quality.

\begin{figure*}[tb]
\centerline{\includegraphics[width=\linewidth]{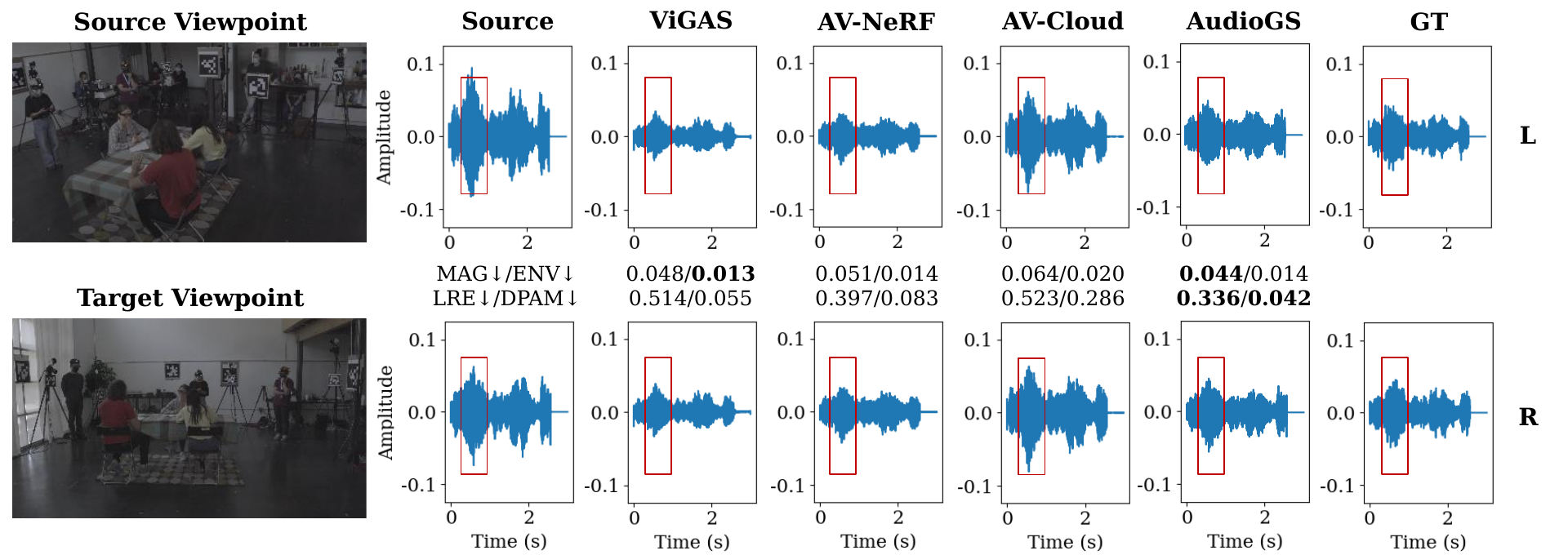}}
\caption{\textbf{Comparison of binaural waveform synthesis.}
\textbf{Left:} Images of source and target viewpoints provided for spatial context. Note that unlike visual-dependent baselines, AudioGS does not utilize these RGB images.
\textbf{Right:} Synthesized waveforms compared against the Ground Truth (GT).
The red boxes highlight regions with distinct transient acoustic events.
Baseline methods (ViGAS, AV-NeRF) often produce attenuated or over-smoothed responses in these high-frequency regions.
In contrast, AudioGS accurately reconstructs the signal envelope, matching the GT closely.
The embedded table reports the scene-specific metrics, where AudioGS achieves the best spatial accuracy (LRE) and perceptual quality (DPAM).}
\label{fig:Visualization}
\end{figure*}

\subsection{Results and Analysis}

Table~\ref{tab:comparison} presents the quantitative comparison on the Replay-NVAS dataset. Our method, AudioGS, achieves the best performance across most metrics in the overall evaluation.

\noindent\textbf{Reconstruction Quality.}
In terms of signal fidelity, AudioGS consistently outperforms baseline methods. Specifically, we reduce the overall MAG by over 14\% and ENV by approximately 6\% compared to the second-best performing methods. This significant margin indicates that our explicit point-based representation preserves the fine-grained time-frequency structure better than implicit neural representations like ViGAS and AV-NeRF, which often suffer from over-smoothing artifacts. Similarly, the point-based method AV-Cloud relies on SfM for geometry initialization, which struggles in texture-less regions and leads to less accurate audio field coverage.

\noindent\textbf{Perceptual and Spatial Fidelity.}
For perceptual quality, AudioGS improves DPAM by 25\% relative to the runner-up, suggesting that our synthesized audio sounds the most realistic and closest to the ground truth to human ears.
Regarding spatial cues, our method shows competitive performance in LRE. AudioGS successfully offers a balanced trade-off between spatial accuracy and audio quality. Overall, AudioGS provides a robust and high-fidelity solution for novel-view acoustic synthesis without relying on visual priors.

\begin{table}[tb]
\centering
\caption{\textsc{Ablation study of key components in AudioGS.}}
\label{tab:ablation}
\begin{tabular}{l|cccc}
\toprule
Methods & MAG $\downarrow$ & ENV $\downarrow$ & LRE $\downarrow$ & DPAM $\downarrow$ \\
\midrule
AudioGS & \textbf{0.0892} & \textbf{0.0300} & \textbf{0.5269} & \textbf{0.0887} \\
w/o DA        & 0.1018 & 0.0341 & 0.8618 & 0.0930 \\
w/o SH         & 0.0911 & 0.0309 & 0.6437 & 0.0896 \\
w/o PC         & 0.0906 & 0.0302 & 0.7932 & 0.0893 \\
\bottomrule
\end{tabular}
\end{table}

\noindent\textbf{Ablation Study.}
To validate the effectiveness of our explicit physical modeling, we remove key components individually. The results in Table~\ref{tab:ablation} demonstrate their specific contributions:
\textbf{1) Distance Attenuation (DA).} DA models the distance-based magnitude decay. Removing this component causes the most severe degradation in spatial accuracy. This confirms that explicitly encoding the physical energy attenuation law is fundamental for establishing correct relative energy levels between ears.
\textbf{2) Spherical Harmonics (SH).}
SH is responsible for encoding directional magnitude radiation. Removing SH leads to a noticeable drop in reconstruction quality. This indicates that SH effectively captures directional energy distribution, refining the ILD cues required for realistic binauralization.
\textbf{3) Phase Correction (PC).}
PC accounts for the wave propagation delay based on geometry. Excluding this module results in phase misalignment, as evidenced by the drop in LRE. This validates that our geometry-guided correction successfully reconstructs ITD cues, which are critical for horizontal localization but cannot be recovered by magnitude modeling alone.

\subsection{Subjective and Qualitative Analysis}
\label{subsec:visualization}

\noindent\textbf{Waveform Reconstruction Quality.}
Fig.~\ref{fig:Visualization} presents a detailed comparison of synthesized waveforms along with source audio and GT.
As observed, implicit methods such as ViGAS and AV-NeRF struggle to capture the full amplitude of these transients, resulting in over-smoothed envelopes and a loss of acoustic energy.
While AV-Cloud captures some high-frequency content, its waveforms differ significantly from GT.
In contrast, AudioGS faithfully reconstructs the fine-grained temporal variations, showing high alignment with the GT within the highlighted regions.

\begin{figure}[tb]
\centerline{\includegraphics[width=\linewidth]{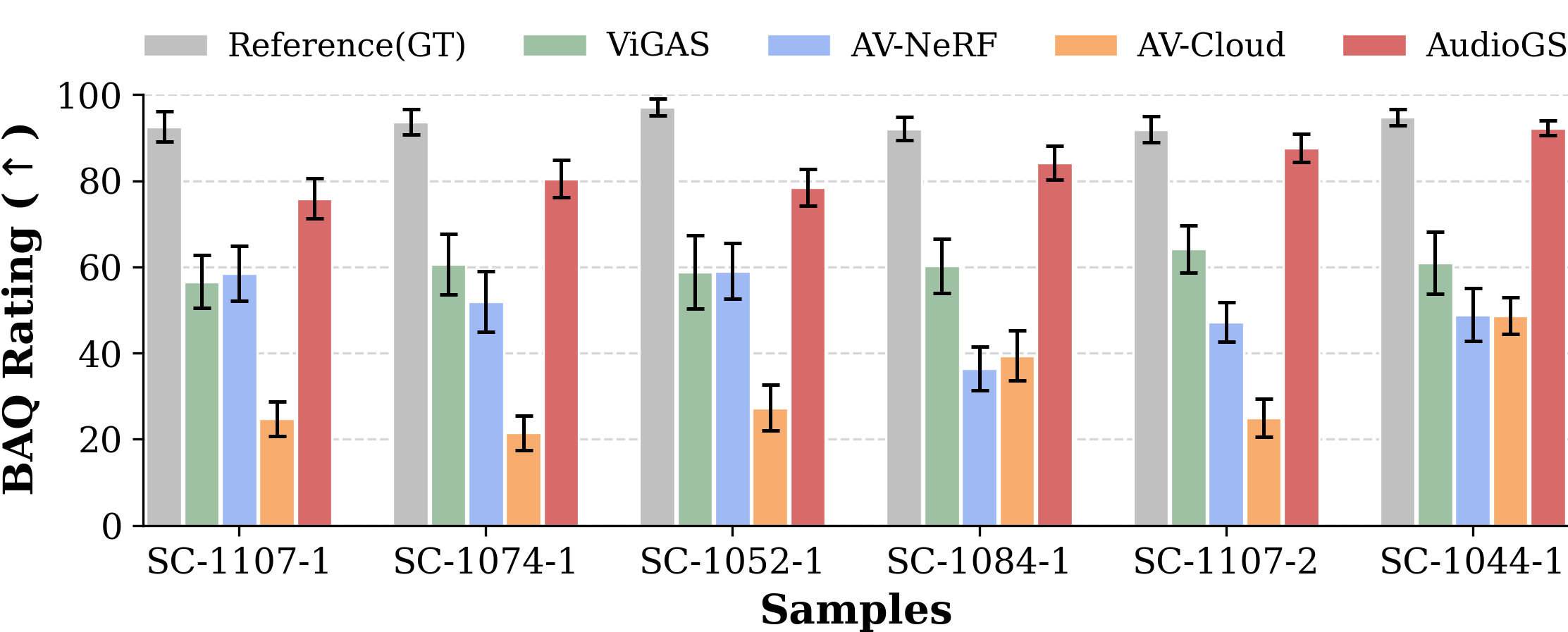}} 
\caption{\textbf{MUSHRA listening test on Replay-NVAS dataset.}
Mean Basic Audio Quality (BAQ) ratings (0--100) from 12 participants on six representative novel-view samples. Error bars denote the standard error of the mean (SEM).}
\label{fig:mushra}
\end{figure}

\begin{figure}[tb]
\centerline{\includegraphics[width=\linewidth]{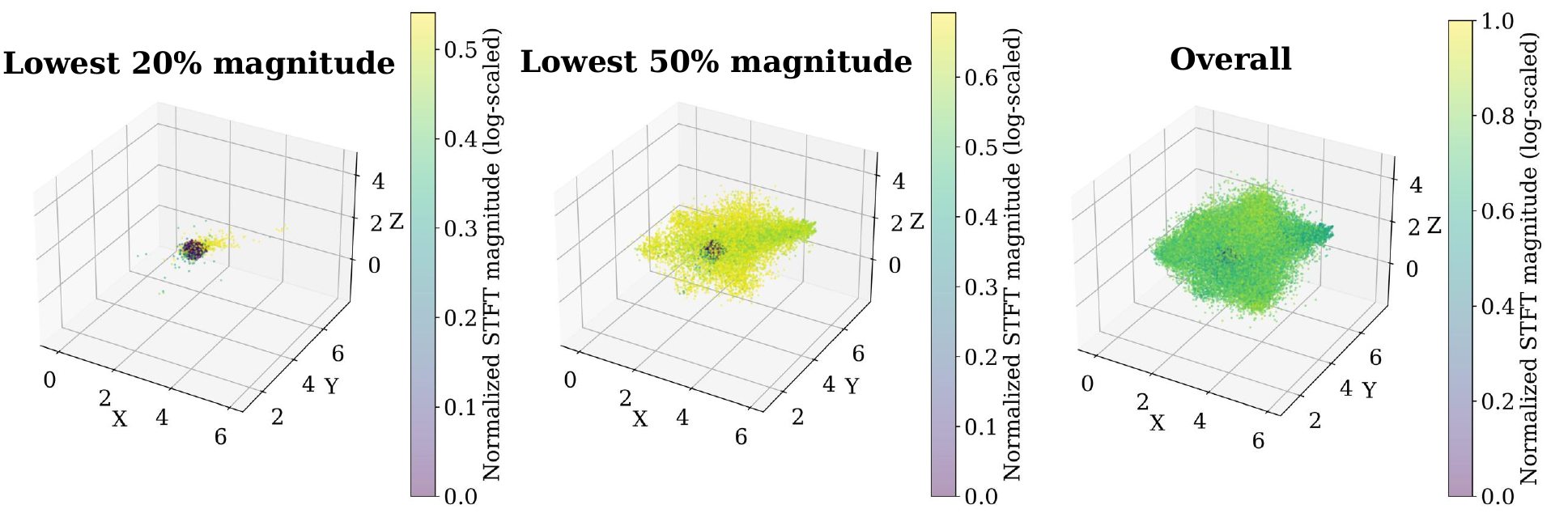}} 
\caption{The spatial distribution of Audio Gaussians filtered by STFT magnitudes. \textbf{Left/Middle:} Gaussians with the lowest 20\% and 50\% magnitudes remain clustered near the initialization center, representing silence or noise. \textbf{Right:} Overall distribution shows that high-energy Gaussians spread out spatially, effectively encoding the directional sound field structure.}
\label{fig:cloud}
\end{figure}

\noindent\textbf{Subjective Listening Test.}
We conduct a MUSHRA listening test with 12 participants~\cite{schoeffler2018webmushra} (students and faculty, including participants with professional audio expertise).
Each participant rates six representative novel-view clips with the ground-truth binaural audio as the reference.
As shown in Fig.~\ref{fig:mushra}, AudioGS achieves the highest mean rating across all samples, consistent with the objective metrics (DPAM) in Table~\ref{tab:comparison}.

\noindent\textbf{Sound Field Structure.}
In Fig.~\ref{fig:cloud}, low-energy Gaussians stay near the initialization center, whereas high-energy Gaussians spread out spatially. This distribution suggests the optimization process effectively identifies the acoustic components to capture spatial cues, while keeping non-informative components compact. This demonstrates AudioGS successfully learns a meaningful spatial representation of the sound field.

\section{Conclusion} \label{sec:conclusion}
We introduced AudioGS, a novel framework for novel-view acoustic synthesis. Unlike previous methods relying on implicit neural fields or visual priors, AudioGS explicitly reconstructs the sound field via a set of learnable Audio Gaussians directly mapped from the spectrogram. By integrating geometry-guided magnitude modulation and phase correction, our model achieves high-fidelity spatial audio rendering. Experiments on the real-world dataset demonstrate that AudioGS significantly outperforms state-of-the-art baselines.
Future work includes extending AudioGS to dynamic scenes via 4D Gaussian Splatting and integrating it with visual 3DGS for unified volumetric audio-visual streaming.
\bibliographystyle{IEEEbib}
\bibliography{icme2026references}
\end{document}